\documentclass[11pt]{article}
\pdfoutput=1
\usepackage{dcolumn}
\usepackage{bm}
\usepackage{graphicx}
\graphicspath{{figs/}}
\usepackage{amssymb,amsmath}
\usepackage{multirow}
\usepackage{cite,color,url}
\usepackage[colorlinks=true
,urlcolor=blue
,anchorcolor=blue
,citecolor=blue
,filecolor=blue
,linkcolor=blue
,menucolor=blue
,linktocpage=true
,pdfproducer=medialab
,pdfa=true
]{hyperref}

\usepackage{slashed}
\usepackage{epsfig,psfrag,rotating,soul}
\usepackage{rotfloat}
\usepackage[font={small}]{caption}
\usepackage{colortbl}

\evensidemargin \oddsidemargin
\allowdisplaybreaks

\let\OLDthebibliography\thebibliography
\renewcommand\thebibliography[1]{
  \OLDthebibliography{#1}
  \setlength{\parskip}{0pt}
  \setlength{\itemsep}{0pt plus 0.3ex}
}

\definecolor{mygray}{gray}{0.85} 
\definecolor{myblue}{cmyk}{0.65, 0.37, 0.0, 0.19}

\begin{document}
\thispagestyle{empty}

\def\thefootnote{\fnsymbol{footnote}}

\begin{flushright}
IFT-UAM/CSIC-21-79
\end{flushright}

\vspace*{1cm}

\begin{center}

\begin{Large}
\textbf{\textsc{Search strategy for gluinos at the LHC \\[.25em] with a Higgs boson decaying into tau leptons}}
\end{Large}

\vspace{1cm}

{\sc
Ernesto~Arganda$^{1, 2}$%
\footnote{{\tt \href{mailto:ernesto.arganda@csic.es}{ernesto.arganda@csic.es}}}%
, Antonio~Delgado$^{3}$%
\footnote{{\tt \href{mailto:adelgad2@nd.edu}{adelgad2@nd.edu}}}%
, Roberto~A.~Morales$^{1}$%
\footnote{{\tt \href{mailto:robertoa.morales@uam.es}{robertoa.morales@uam.es}}}%
, and Mariano~Quir\'os$^{4}$%
\footnote{{\tt \href{quiros@ifae.es}{quiros@ifae.es}}}%

}

\vspace*{.7cm}

{\sl
$^1$Instituto de F\'{\i}sica Te\'orica UAM/CSIC, \\
C/ Nicol\'as Cabrera 13-15, Campus de Cantoblanco, 28049, Madrid, Spain

\vspace*{0.1cm}

$^2$IFLP, CONICET - Dpto. de F\'{\i}sica, Universidad Nacional de La Plata, \\ 
C.C. 67, 1900 La Plata, Argentina

\vspace*{0.1cm}

$^3$Department of Physics, University of Notre Dame, 225 Nieuwland Hall \\
Notre Dame, IN 46556, USA

\vspace*{0.1cm}

$^4$Institut de F\'{\i}sica d'Altes Energies (IFAE) and BIST, Campus UAB \\
08193, Bellaterra, Barcelona, Spain

}

\end{center}

\vspace{0.1cm}

\begin{abstract}
\noindent
The possibility in supersymmetric scenarios that the dark matter candidate is a Higgsino-like neutralino means that its production can be associated with Higgs bosons. Taking advantage of this fact, we propose a LHC search strategy for gluinos with $\tau$ leptons in the final state, coming from the decay of a Higgs boson. We consider the strong production of a pair of gluinos, one of which decays into the Higgsino plus jets while the other decays into the bino plus jets. In turn, this bino decays into the Higgsino plus a Higgs boson which finally decays into a $\tau$-lepton pair. Therefore, the experimental signature under study consists of 4 jets, 2 $\tau$ leptons, and a large amount of missing transverse energy. Our cut-based search strategy allows us to reach, for a LHC center-of-mass energy of 14 TeV and a total integrated luminosity of 1 ab$^{-1}$, significances of up to 2 standard deviations, considering systematic uncertainties in the SM background of 30\%. The projections for 3 ab$^{-1}$ are encouraging, with significances at the evidence level, which in more optimistic experimental scenarios could exceed 4 standard deviations.
\end{abstract}

\def\thefootnote{\arabic{footnote}}
\setcounter{page}{0}
\setcounter{footnote}{0}

\newpage

\section{Introduction}
\label{intro}
As the Standard Model (SM) naturalness problem is linked to the very existence of the Higgs boson, its experimental discovery at the Large Hadron Collider (LHC)~\cite{Aad:2012tfa,Chatrchyan:2012ufa} led to the strong belief that beyond the SM (BSM) physics should be somehow related to the Higgs sector. This belief motivated considering supersymmetry (SUSY), and in particular the minimal supersymmetric extension of the SM (MSSM)~\cite{Nilles:1983ge,Haber:1984rc,Gunion:1984yn}, as a possible ultraviolet completion where the naturalness problem is solved. Moreover, as an interesting spinoff, the MSSM (in the presence of $R$-parity symmetry which protects proton decay) was shown to contain a candidate for dark matter (DM), the lightest supersymmetric particle (LSP), most likely the lightest neutralino. As the LSP is stable, its presence should be ``detected" as missing energy ($E_T^\text{miss}$), which is the characteristic signature of supersymmetric searches. 

After a plethora of direct DM searches, the case of the LSP being identified as a (nearly) pure Higgsino $\tilde\chi^0_1$ (the SUSY partner of the Higgs boson) with a mass of $\sim 1.2$ TeV, remains as a viable and exciting possibility~\cite{Kowalska:2018toh}, as many of its properties are inherited from the Higgs sector and are so rooted in the solution of the naturalness problem. This possibility was thoroughly studied in Ref.~\cite{Delgado:2020url} where imposing correct electroweak breaking, the correct Higgs mass and some unification properties of supersymmetric parameters at the unification scale $M_U\sim 2\times 10^{16}$ GeV, the supersymmetric mass space was constrained, in particular in the gaugino/Higgsino sector. 

As the production cross section of the Higgsino is too low for discovery at the LHC, an alternative search is gluino $\tilde g$ production followed by a decay chain leading to the LSP. As the standard searches use the gluino decay channel being (100\%) $\tilde g\to \tilde\chi^0_1jj$, we have proposed in a recent paper the case where the gluino entirely decays as $\tilde g\to \tilde\chi^0_3jj$ (where $\tilde\chi_3^0$ is a heavier (nearly) bino) followed by the decay $\tilde\chi_3^0\to \tilde\chi_1^0 h$, and so we were studying the signal $pp\to\tilde g\tilde g$ $\to$ $4j+4b+E_T^\text{miss}$~\cite{Arganda:2021lpg}. This possibility appears in models where the first and second squark generation are heavy and degenerate in mass while the third-generation squarks are heavier and thus in practice fully decoupled, a possibility which can appear in the effective theory of superstring models~\cite{Brignole:1997wnc}.  

However, depending on the supersymmetric squark spectrum, the gluino can have non-vanishing branching ratios for the channel $\tilde g\to \tilde \chi_1^0jj$ and $\tilde g\to\tilde\chi_3^0jj$ and so in this paper we will consider the case of asymmetric decays where one of the produced gluinos decays into $\tilde\chi_1^0jj$ and the other into $\tilde\chi_1^0hjj$ and, to partly avoid the QCD background, we will consider the decay of the SM-like Higgs bosons into a $\tau$-lepton pair, $h\to\tau^+\tau^-$. 

The rest of the paper is organized as follows: In Section~\ref{th-frame} we summarize the theoretical framework of the MSSM scenarios with Higgsino-like dark matter that we work with. Section~\ref{collider} is devoted to the collider analysis, in which we characterize the signal against the SM backgrounds, develop our search strategy and show the signal significances one can expect in the high-luminosity phase of the LHC. Finally, we leave Section~\ref{conclus} for conclusions and final remarks.

\section{Theoretical Framework}
\label{th-frame}

As already explained we will consider the same spectrum as in our previous work~\cite{Arganda:2021lpg} with the lightest electroweakinos ($\tilde\chi_{1,2}^0$, $\tilde\chi_1^\pm$) being mostly Higgsino-like, nearly degenerate with a mass of $\sim$1.2 TeV, a bino-like neutralino ($\tilde\chi_3^0$) with a mass of $\sim$1.5 TeV and a gluino ($\tilde g$) with a mass of $\sim$1.7 TeV. The rest of the supersymmetric spectrum is heavy and provides the appropriate branching ratios we have considered in the numerical analysis.

As stated in  the previous section, the gluino can then either decay to the Higgsino-like doublet plus jets or to the bino-like singlet plus jets, the relative branching ratios are controlled by the spectrum of squarks. The bino will decay to the LSP and the SM-like Higgs boson. The fact that one can produce Higgs bosons in chains is a generic feature of any spectrum where the LSP is mostly Higgsino. This opens up the possibility of exploring discovery channels for this kind of spectrum in different channels of Higgs boson decays. One of these possibilities is when the Higgs bosons decay into a $\tau$-lepton pair, which is the case we will consider in this paper.

It is important to notice here that supersymmetric searches by ATLAS~\cite{ATLAS:2017qwn,ATLAS:2018diz} and CMS~\cite{CMS:2019eln} with taus in the final state are based on models with light staus, which are produced in colliders, directly ($\tilde\tau\to\tau\tilde\chi_1^0$) or in chargino/neutralino decays ($\tilde\chi_1^{0}\to \tilde\tau\tau$, $\tilde\chi_1^\pm\to\tau\tilde\nu_\tau$), and the produced $\tau$ leptons are related to the missing transverse energy. In our case we are producing strongly-coupled gluinos whose production is enhanced relative to the former. Moreover our $\tau$-lepton pairs are unrelated to the $E_T^\text{miss}$ but they stem from Higgs decays so that its invariant mass can reproduce the Higgs mass. So our proposed channel with a $\tau$-lepton pair might be disentangled from the present supersymmetric searches with taus in the final state. 

\section{Collider Analysis}
\label{collider}

\begin{figure}[h!]
	\begin{center}
		\begin{tabular}{c}
			\centering
			\includegraphics[scale=0.4]{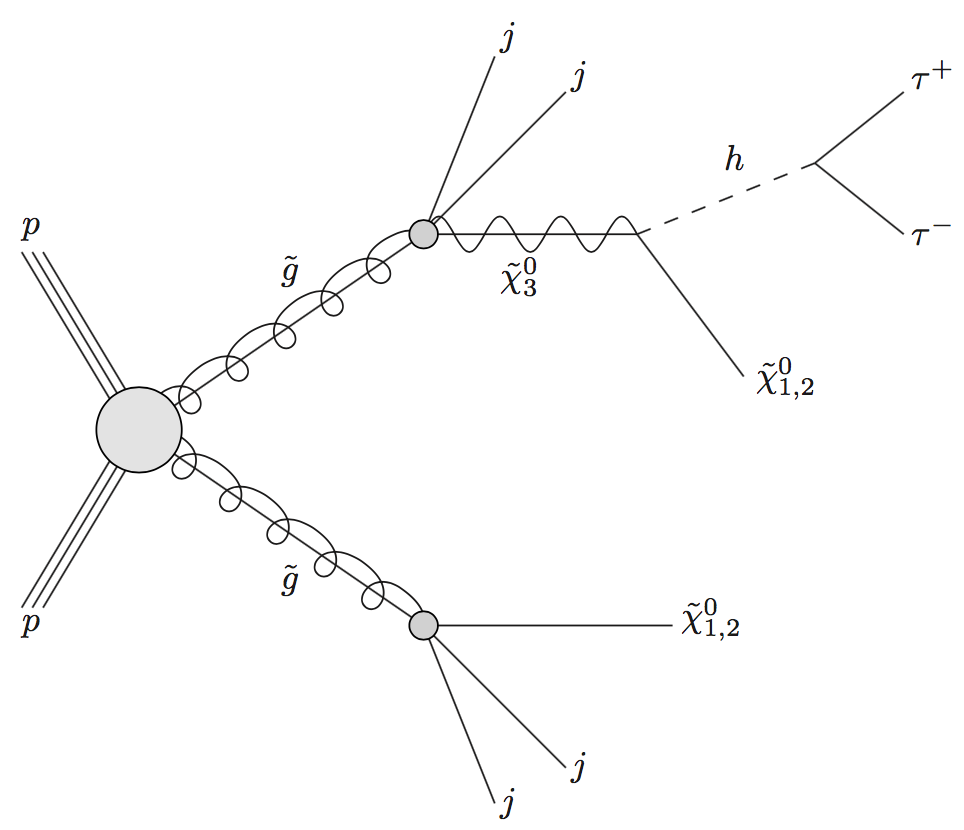}
		\end{tabular}
		\caption{\it Gluino decay chain relevant for this work.}
		\label{gluino-cascade}
	\end{center}
\end{figure}

In this work we study an asymmetric SUSY decay chain at the LHC that comes from the production of a gluino pair, $pp \to \tilde g \tilde g$ as in Fig.~\ref{gluino-cascade}. One of the gluinos decays directly into the LSP Higgsino plus two light jets ($\tilde g \to \tilde \chi_{1,2}^0 j j$). However, the other gluino decays into $\tilde \chi_3^0$ and two light jets ($\tilde g \to \tilde \chi_3^0 j j$), while $\tilde \chi_3^0$ goes to $\tilde \chi_{1,2}^0$ and the 125-GeV SM-like Higgs boson, which in turn decays into a $\tau$-lepton pair. Thus, the experimental signature under study consists of four light jets, a hadronic $\tau$-lepton pair, and a large amount of missing transverse energy ($4 j+ 2 \tau + E_T^\text{miss}$), whose main backgrounds are QCD multijet, $Z$ + jets and $W$ + jets productions; $t \bar t$ production; $t \bar t$ production in association with Higgs and electroweak bosons, $t \bar t$ + $X$ ($X$ = $h$, $W$, $Z$); and diboson production ($WW$, $ZZ$, $WZ$) plus jets.

The LHC search strategy is developed for a center-of-mass energy of $\sqrt{s}$ = 14 TeV and a total integrated luminosity of $\cal{L}$ = 1 ab$^{-1}$, corresponding with the high-luminosity LHC (HL-LHC) phase. Both signal and background events are generated with {\tt MadGraph\_aMC@NLO 2.8.1}~\cite{Alwall:2014hca}, whilst we make use of {\tt PYTHIA 8.2}~\cite{Sjostrand:2014zea} for the parton showering and hadronization and {\tt Delphes 3.3.3}~\cite{deFavereau:2013fsa} for the simulation of the detector response.

Therefore, in order to reduce the large background cross sections and make Monte Carlo event simulation more efficient, the following generator-level cuts on the $p_T$ of the light jets and $\tau$ leptons for the background are imposed~\footnote{For the signal event generation, the default cuts on the $p_T$ of the light jets and $\tau$ leptons have been used ($p_T^j >$ 20 GeV and $p_T^\tau >$ 20 GeV).}:
\begin{eqnarray}
p_T^{j_1} > 150 \, \text{GeV} \,, \quad p_T^{j_2} > 80 \, \text{GeV} \,, \quad p_T^{j_3} > 20 \, \text{GeV} \,, \quad p_T^{j_4} > 20 \, \text{GeV} \,, \nonumber\\
p_T^{\tau_1} > 20 \, \text{GeV} \,, \quad p_T^{\tau_2} > 20 \, \text{GeV} \,,
\label{generatorcuts}
\end{eqnarray}
where $j_1 - j_4$ ($\tau_1 - \tau_2$) runs from the most to the least energetic light jet ($\tau$ lepton). In order to deal with the many jets in the final state, we have implemented the MLM algorithm~\cite{Mangano:2002,Mangano:2006rw} for jet matching and merging. We set {\tt xqcut} to 20 for all generated samples and {\tt qcut} equal to 550, 50, and 30 for signal, $t \bar t$-like and backgrounds with bosons, respectively, to optimize the simulation and check that the distributions of the related jets are smooth. In addition, we perform the simulation using a working point for the $\tau$-tagging efficiency of $\epsilon_\tau$=90\%~\cite{ATLAS:2017mpa,Sirunyan:2018pgf} and a misidentification-rate equal to 0.02. The simulation input files and the internal analysis codes are available upon request to the authors.

By means of {\tt SOFTSUSY 4.1.10}~\cite{Allanach:2001kg,Allanach:2017hcf,Allanach:2013kza,Allanach:2009bv,Allanach:2011de,Allanach:2014nba,Allanach:2016rxd} we compute the SUSY spectrum and branching ratios for our benchmark, and we obtain the value of the gluino-pair production cross section from~\cite{LHCSUSYxs}. The masses of the particles involved in the proposed decay chain are as follows: $M_{\tilde g}$ = 1.7 TeV, $m_{\tilde \chi_3^0}$ = 1.5 TeV, and $m_{\tilde \chi_1^0}$ = 1.2 TeV. The production cross section of two gluinos of 1.7 TeV is $\sigma(pp \to \tilde g \tilde g)$ = 7.6 fb, and the branching ratios of our SUSY decay chain are BR($\tilde g \to \tilde \chi_1^0 j j$) = 0.39, BR($\tilde g \to \tilde \chi_3^0 j j$) = 0.42, BR($\tilde \chi_3^0 \to \tilde \chi_1^0 h$) = 0.27, and BR($h \to \tau^+\tau^-$) = 0.06. Notice that the decay of the third neutralino $\tilde\chi_3^0$ to the other (almost degenerated) LSP is negligible. Then, for a luminosity of $\cal{L}$ = 1 ab$^{-1}$, 19.6 signal events are expected.

Concerning the backgrounds, it is important to mention the following issues:
\begin{itemize}
\item QCD multijet. This background is relevant when jets are misidentified as $\tau$ leptons and large missing transverse momentum is induced by jet energy mismeasurements. In general it is treated with data-driven techniques and is beyond our computational capacity. The fact of demanding a large amount of $E_T^\text{miss}$ allows us to reject the instrumental missing energy related to this background, by means of the $E_T^\text{miss}$ significance and spatial configurations, as it is usual in the experimental searches. In particular, we follow \cite{Aaboud:2018mna} in order to give an estimation of this background since it corresponds to the same signature and a similar strategy was implemented. 
\item Diboson + jets. The $WW$+jets and $ZZ$+jets productions are the main backgrounds in this category. We simulated up to two extra jets. Taking into account our parton level cuts and BR($W\to\tau\nu$) = 0.11, BR($Z\to\nu\nu$) = 0.21, and BR($Z\to\tau\tau$) = 0.03, we expect $1\times 10^{4}$ events for $WW$+jets and $1.2\times 10^{3}$ for $ZZ$+jets at ${\cal L}=1000$ fb$^{-1}$.
\item $t\bar{t}$ production. This process with both top quarks decaying into $\tau$ leptons is the most dangerous background, in spite of BR($t\bar{t}_{\rm tau}$) = 0.01. We expect $7.1\times 10^{5}$ events after the generator cuts and the inclusion of one extra jet in the simulation results in $6.3\times 10^{5}$ events more. Furthermore, we consider an estimation of $t\bar{t}$+2$j$ taking into account an extra 10\% factor to the simulated $t\bar{t}_{\rm tau}$ and $t\bar{t}_{\rm tau}$+$j$ events (given by the ratio of the corresponding cross sections).
\item $t\bar{t} + X$ backgrounds. Related to the previous background, we consider the $t\bar{t}_{\rm semitau}+W(\tau\nu)$ case with BR($t\bar{t}_{\rm semitau}$) = 0.15 and $t\bar{t}_{\rm had}+Z(\tau\tau)$, $t\bar{t}_{\rm had}+h(\tau\tau)$ cases with BR($t\bar{t}_{\rm had}$) = 0.46 Then we expect $1.1\times 10^{3}$ events in this category for our generation setup.
\item $V$+jets production: we consider $W(\tau\nu)+3j$ and $Z(\tau\tau)+3j$ as the main reducible backgrounds in this category. In the first case, a second $\tau$ lepton can arise from a jet misidentified as a fake $\tau$ lepton. In the second case, the jet energy mismeasurements produce the large missing energy. Then we expect $1.76\times 10^{5}$ and $1.8\times 10^{5}$ events at ${\cal L}$ = 1000 fb$^{-1}$ with our generation setup, respectively.
\end{itemize}

\begin{figure}
	\begin{center}
		\begin{tabular}{c}
			\centering
			\hspace*{-3mm}
			\includegraphics[scale=0.5]{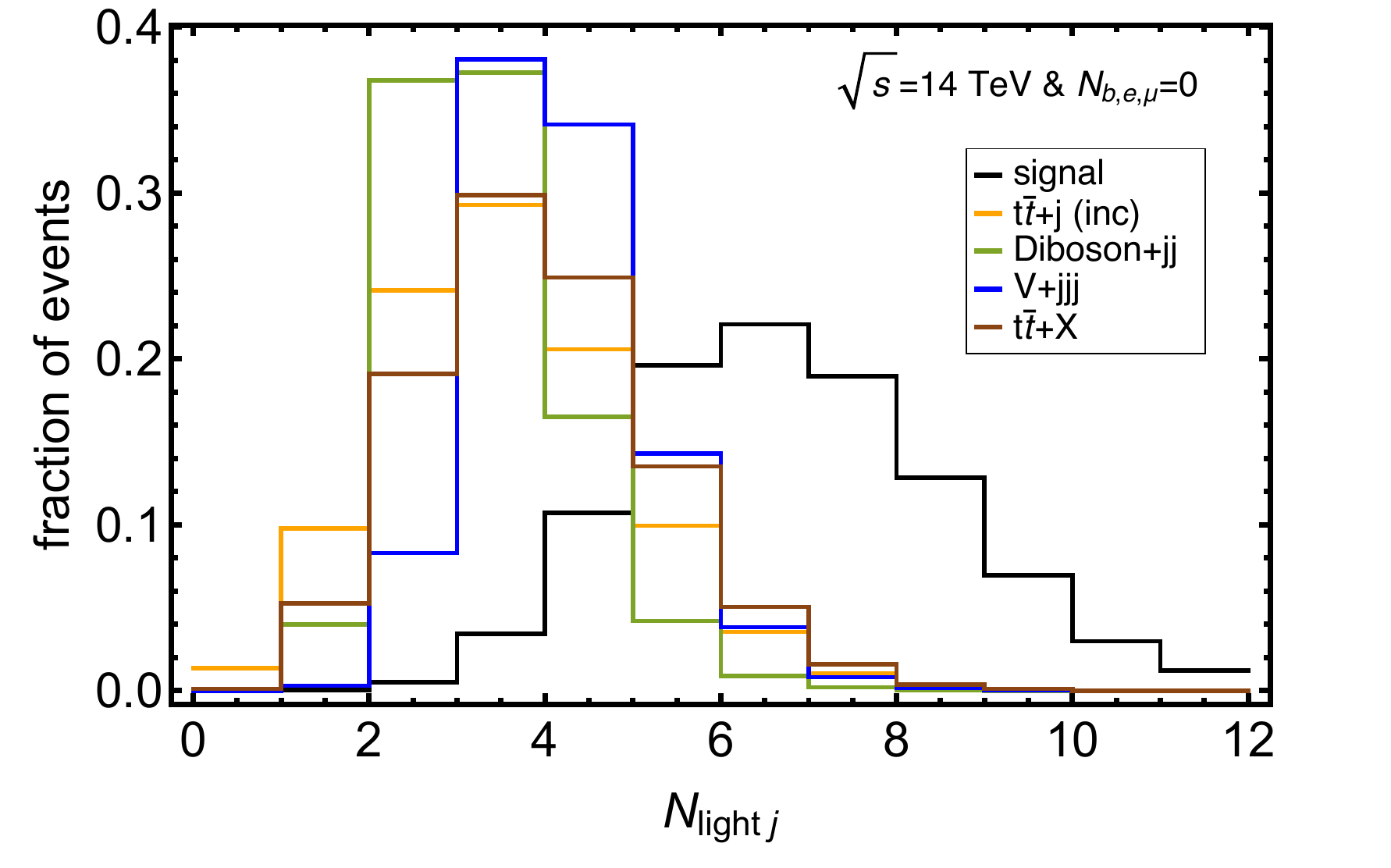}
		\end{tabular}
		\caption{\it Distributions of the fraction of signal and background events corresponding to the number of tagged light jets $N_j$.}
		\label{fig:Nj}
	\end{center}
\end{figure}

We will demand exactly two hadronic $\tau$ leptons in the final state in order to tagging the Higgs boson of our signal of interest. Then this $\tau$-lepton pair comes from a disintegration at the end of the decay chain and it is not closely related to the missing transverse energy or other SUSY intermediate states (as it is usual in sleptons signals). On the other hand, as we do not expect bottom quarks, electrons and muons in our signal, the corresponding vetoes will be imposed in order to characterized the signal against the background. In addition, the distributions of the fractions of events of the number of light jets, $N_j$, is shown in Fig~\ref{fig:Nj}.
With these results in mind, we can define the following selection cuts that characterize our signal:
\begin{eqnarray}
\quad N_\tau = 2 \,, \quad N_j \geq 4 \,, \quad N_{b,e,\mu} = 0 \,.
\label{SRdefinitions}
\end{eqnarray}

\begin{figure}
	\begin{center}
		\begin{tabular}{cc}
			\centering
			\hspace*{-3mm}
			\includegraphics[scale=0.4]{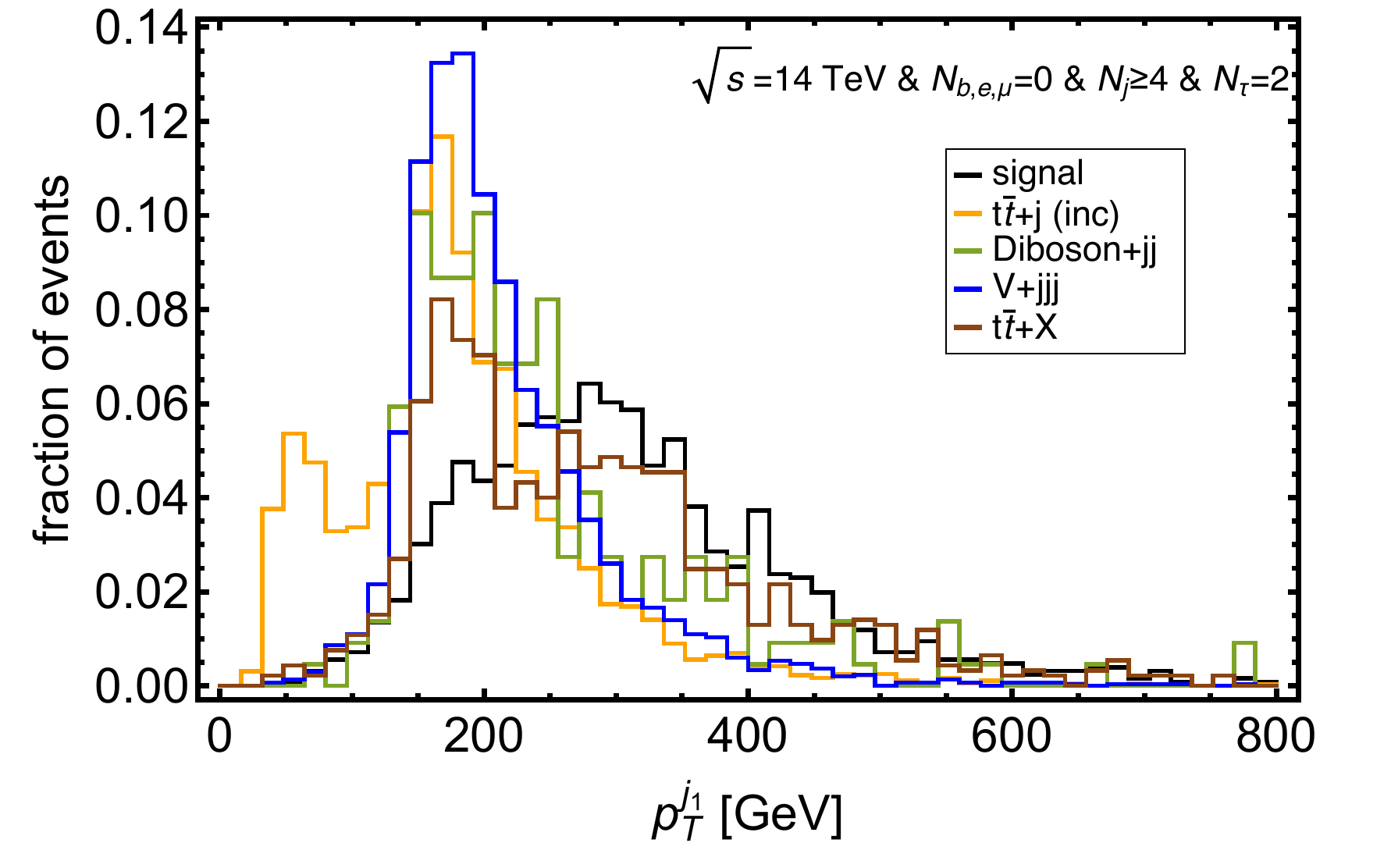} &
			\includegraphics[scale=0.4]{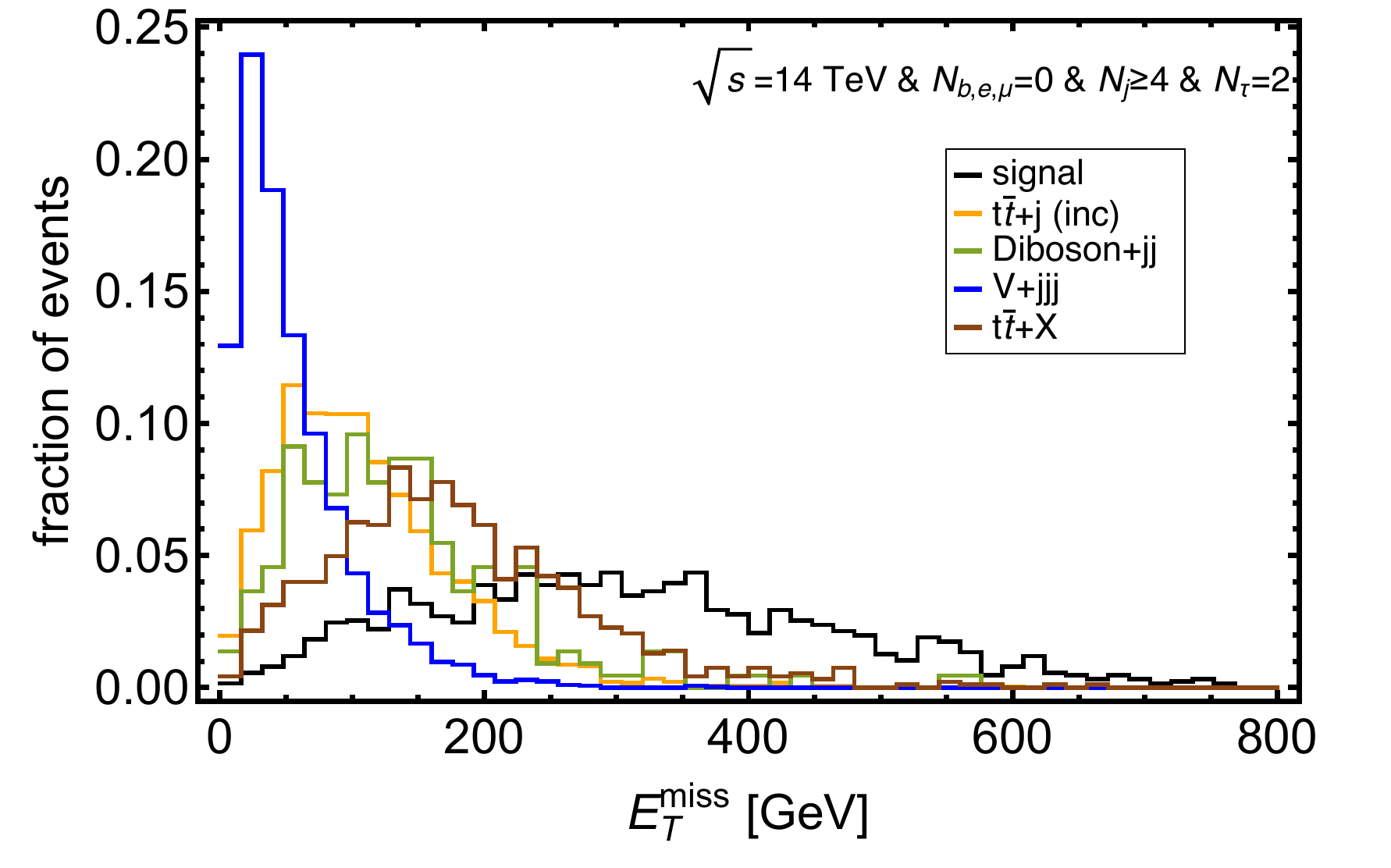} \\
			\includegraphics[scale=0.4]{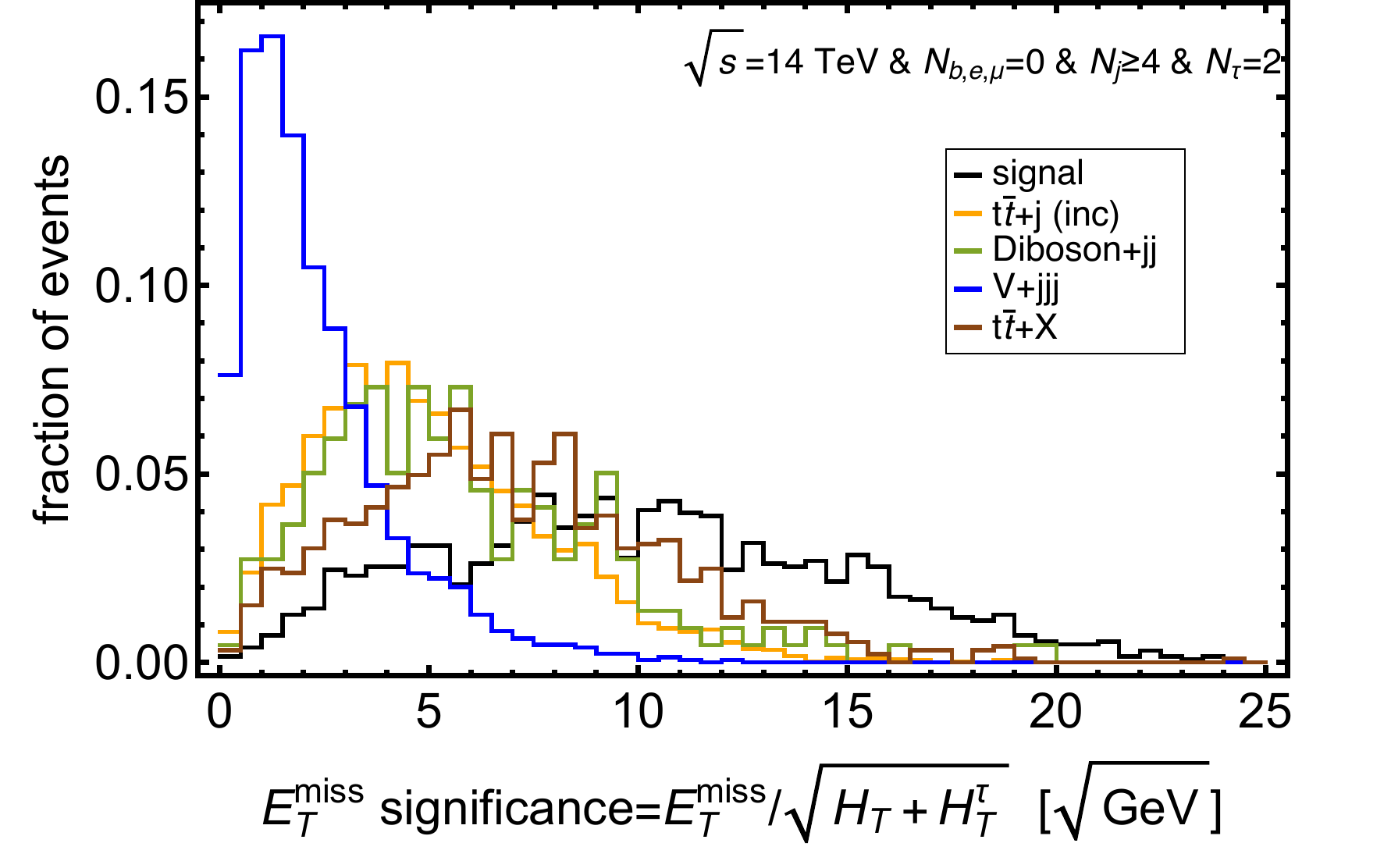} &
			\includegraphics[scale=0.4]{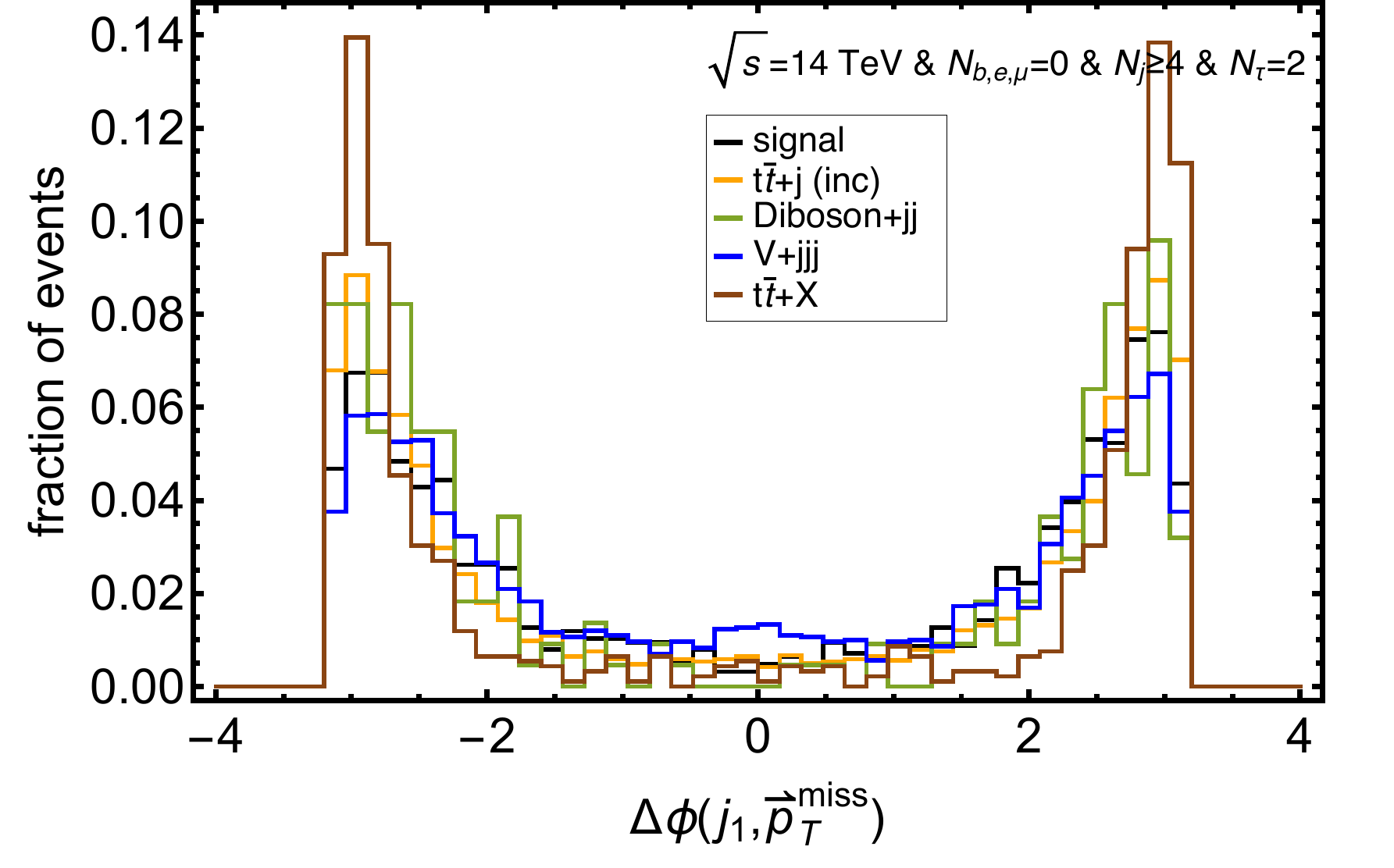} \\
			\includegraphics[scale=0.4]{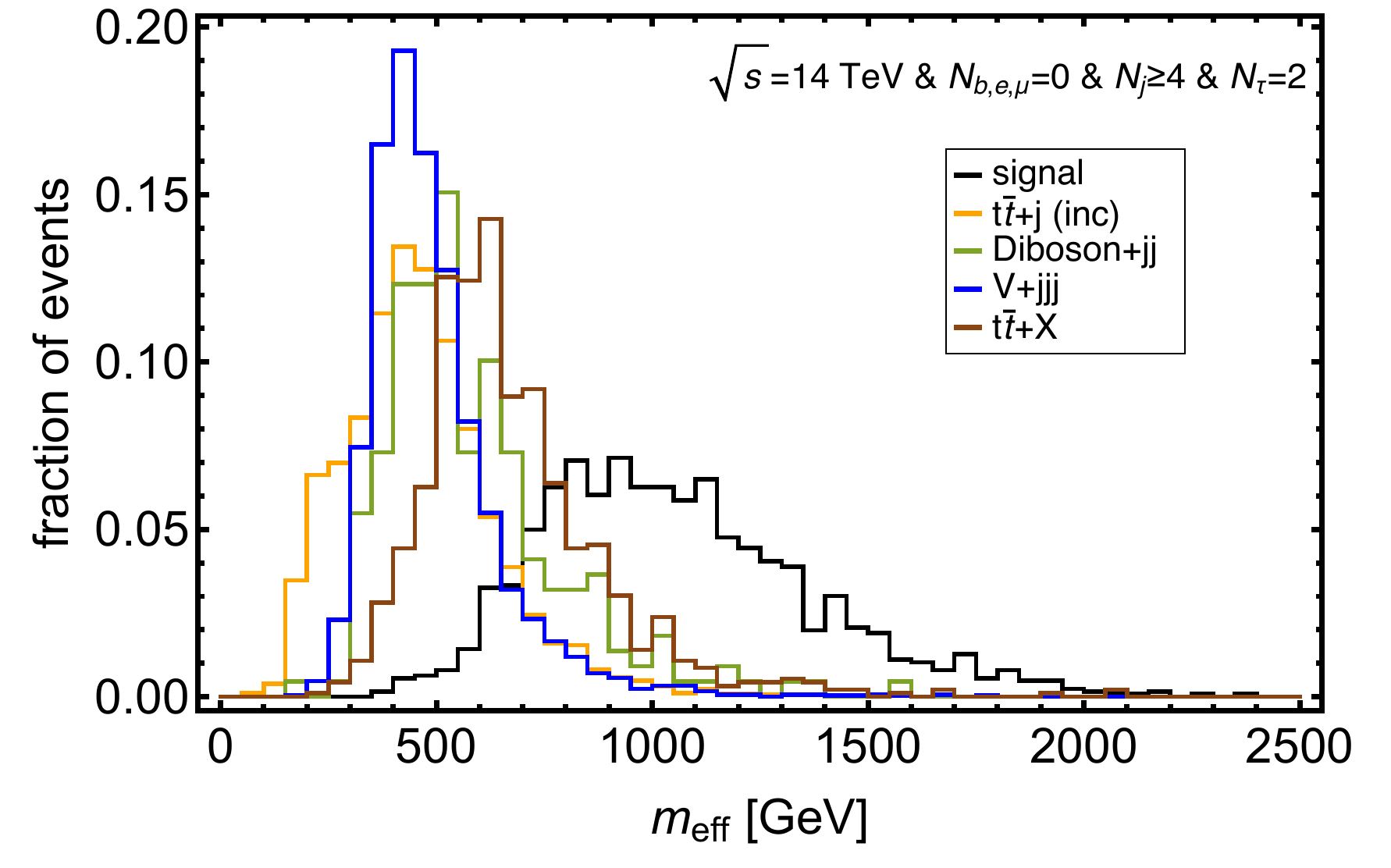} &
			\includegraphics[scale=0.4]{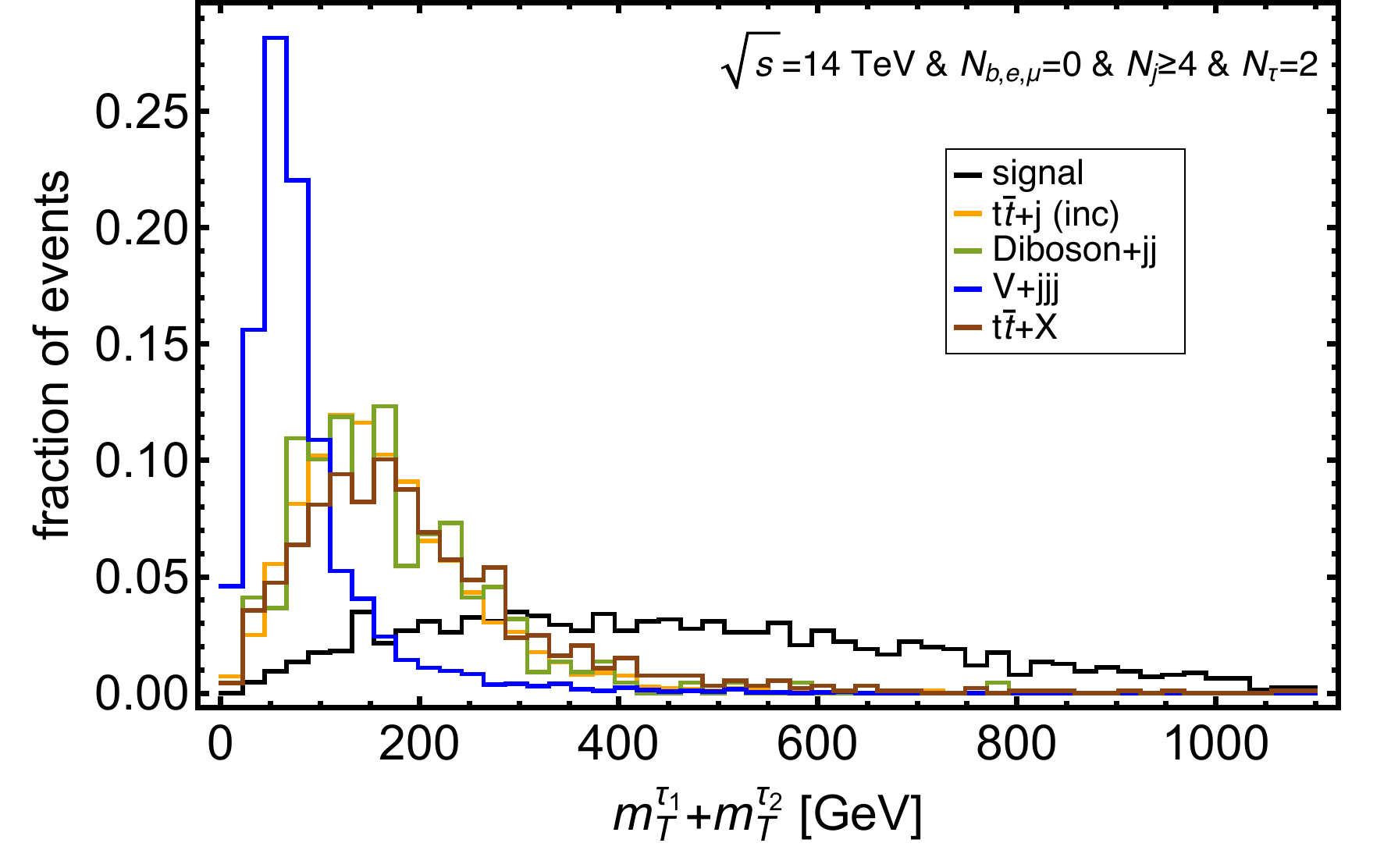}
		\end{tabular}
		\caption{\it Distributions after selection cuts in Eq.~\ref{SRdefinitions} of the fraction of signal and background events of the transverse momentum of the leading light jet $p_T^{j_1}$ (upper left panel), the missing transverse energy $E_T^\text{miss}$ (upper right panel), the $E_T^\text{miss}$ significance (medium left panel), the azimuthal angle difference $\Delta\phi$($j_1$, $\vec{p}_T^\text{\,miss})$ between the leading jet and the $\vec{p}_T^\text{\,miss}$ (medium right panel), the effective mass $m_\text{eff}$ (lower left panel), and the sum of the transverse masses of the $\tau$ leptons (lower right panel).}
		\label{fig:cutflow-plots}
	\end{center}
\end{figure}

We devote Fig.~\ref{fig:cutflow-plots} to six decisive kinematic variables, for which we show the distributions (after the selection cuts) of the fraction of signal and background events: the transverse momentum of the leading light jet $p_T^{j_1}$ (upper left panel); the missing transverse energy, $E_T^\text{miss}$ (upper right panel); the $E_T^\text{miss}$ significance (medium left panel) defined as the ratio of the missing transverse energy over the square root of the hadronic activity $H_T=\sum_{\rm all\,j} p_T$ plus tau activity $H_T^\tau=\sum_{\rm all\,\tau} p_T$; the azimuthal angle difference between the leading jet and the missing transverse momentum $\vec{p}_T^\text{\,miss}$, $\Delta\phi$($j_1$, $\vec{p}_T^\text{\,miss})$ (medium right panel); the effective mass, $m_\text{eff}$ (lower left panel), defined as the sum of $E_T^\text{miss}$ plus the hadronic activity ($m_\text{eff}$ = $E_T^\text{miss} + H_T$); and the sum of the transverse masses of the $\tau$ leptons, $m_T^{\tau_1}$+$m_T^{\tau_2}$ (lower right panel) in which $m_T^\tau=\sqrt{2p_T^\tau E_T^\text{miss}(1-\cos\Delta\phi(\vec{p}^{\,\tau}, \vec{p}_T^\text{\,miss}))}$.

Many interesting features can be obtained from this Fig.~\ref{fig:cutflow-plots}. Starting with the $p_T^{j_1}$ distributions, the corresponding ones to the backgrounds are all chopped at values between 150 GeV and 200 GeV, while the signal presents the highest fraction of events for values above 250 GeV. We will see later, when we will define our search strategy, that a cut on $p_T^{j_1}$ together with another on $p_T^{j_2}$, whose distributions we do not show here for space saving, will be very useful to increase the signal-to-background ratio.

The $E_T^\text{miss}$ distributions of signal and background present different patterns. The latter have their largest event fractions for values below 200 GeV and practically disappear at 400 GeV. On the other hand, the signal distribution is more or less flat in the interval from 200 GeV to 400 GeV, and extends beyond 600 GeV. In that sense, it is to be expected that a strong cut on this variable would be really efficient in killing a large part of the background while preserving an important portion of the signal events.

On the other hand, the $E_T^\text{miss}$ significance distributions for the backgrounds have peaks near 5 GeV$^{1/2}$ while the signal have most of its events above this value. Also, $\Delta\phi$($j_1$, $\vec{p}_T^\text{\,miss})$ distributions are virtually identical for all the considered backgrounds and our signal. 
Standard cuts on these two variables ($E_T^\text{miss}$ significance $>$ 5 GeV$^{1/2}$ and $\vert\Delta\phi$($j_1$, $\vec{p}_T^\text{\,miss})\vert>$ 0.4) do not affect the signal events too much but they reject most of the QCD multijet background, in which poorly measured jets or neutrinos emitted closed to the axis of a jet produce large missing energy. 
Supported by the data-driven analysis of this background in \cite{Aaboud:2018mna} corresponding to the `2$\tau$ channel', we will include 0.3 events at the end of our cut-based analysis for ${\cal L}$=1 ab$^{-1}$. Notice that it is a conservative estimation since it corresponds to a direct extrapolation from ${\cal L}=36.1$ fb$^{-1}$ to the luminosity considered here.

Fortunately, the $m_\text{eff}$ variable is also shown to be really efficient in discriminating signal from background, since all the distributions of the latter have peaks below 700 GeV, while the signal presents a broad peak around 1000 GeV, extending beyond 1500 GeV with a non-negligible fraction of events. 

In addition, the sum of the transverse masses of the two $\tau$ leptons is another very interesting variable. The signal distribution is practically flat in the range from 200 GeV to 1000 GeV, while the distributions of the backgrounds have their peaks of maximum event fraction values below 300 GeV, and from this value they drop strongly, being negligible for values larger than 400 GeV.

Taking into account everything discussed above, we impose the following cuts at detector level, called `MET cuts' in short:
\begin{eqnarray}
&& p_T^{j_1} > 170 \, \text{GeV} \,, \quad p_T^{j_2} > 90 \, \text{GeV} \,, \nonumber\\
&& E_T^\text{miss}>150 \, \text{GeV} \,, \quad \vert\Delta\phi(j_1,\vec{p}_T^\text{\,miss})\vert>0.4 \,,
\label{pTcuts}
\end{eqnarray}
and define our search strategy with the following steps:
\begin{itemize}
\item Selection cuts of Eq.~(\ref{SRdefinitions}),
\item `MET cuts' of Eq.~(\ref{pTcuts}),
\item $m_\text{eff}>$ 1000 GeV,
\item and $m_T^{\tau_1}+m_T^{\tau_2}>$ 450 GeV.
\end{itemize}

From the experimental point of view, the same signature was studied by the ATLAS Collaboration~\cite{Aaboud:2018mna}, and a similar search strategy was developed there. However the spectrum considered in that analysis is very different to ours and the resulting topology of the SUSY decay chains too. In particular, in that search the $\tau$ leptons are associated to the $\tau$ sleptons, $\tilde{\tau}$, and the LSP, while in our case the $\tau$ leptons are related to the Higgs boson. Previous similar experimental searches have been reported by ATLAS~\cite{ATLAS:2012ht,ATLAS:2014eel,ATLAS:2016mqw} and CMS~\cite{CMS:2013pbl}.

To get an idea of the signal significance that we can obtain with this search strategy, we will use, on the one hand, the expression of the statistical significance, defined in~\cite{Cowan:2010js} as:
\begin{equation}
\mathcal{S}_\text{sta}=\sqrt{-2 \left((B+S) \log \left(\frac{B}{B+S}\right)+S\right)} \,,
\label{statS}
\end{equation}
where $S$ represents the number of signal events and $B$ is the number of background events. On the other hand, if we want to take into account our lack of knowledge about the background, we can calculate a more realistic estimate of the signal significance by means of the following expression~\cite{Cowan:2010js}:
\begin{equation}
{\cal S}_\text{sys} = \sqrt{2 \left((B+S) \log \left(\frac{(S+B)(B+\sigma_{B}^{2})}{B^{2}+(S+B)\sigma_{B}^{2}}\right)-\frac{B^{2}}{\sigma_{B}^{2}}\log \left(1+\frac{\sigma_{B}^{2}S}{B(B+\sigma_{B}^{2})} \right) \right)} \,,
\label{systS}
\end{equation}
where $\sigma_{B}=(\Delta B) B$, with $\Delta B$ being the relative systematic uncertainty, chosen here to be of 30\%.~\footnote{Using the {\tt Zstats} package~\cite{Zstats}, we have verified that the resulting ${\cal S}_\text{sta}$ and ${\cal S}_\text{sys}$ are compatible with the values obtained with the expressions for discovery significances proposed in~\cite{Kumar:2015tna,Bhattiprolu:2020mwi}, with differences of 5-10\%.}

\begin{table}
\hspace*{-7.5mm}
    \centering
\begin{tabular}{r|rrrrr|cc}
\hline\hline
   Process  & Signal & $t\bar{t}+2j$ (inc.) & $t\bar{t}+X$ & Diboson & $V$+jets & $\cal{S}_\text{sta}$ & $\cal{S}_\text{sys}$ \\
   \hline
    Expected  & 19.6 & $1.47 \times 10^6$ & $1.1 \times 10^3$ & $1.12 \times 10^4$ & $3.56\times 10^5$ & $0.01$ & $4\times 10^{-5}$ \\
    \hline
    Selection cuts  & 5.6 & 3283.7 & 9.9 & 167.5 & 9952.4 & 0.05 & $1.4\times 10^{-3}$ \\
    `MET cuts'  & 3.82 & 772.3 & 2.31 & 44.3 & 262.7 & 0.12 & $1.2\times 10^{-2}$ \\
    $m_\text{eff}$ $>$ 1000 GeV  & 2.6 & 9.8 & 0.4 & 7.9 & 43.8 & 0.33 & 0.13 \\
    $m_T^{\tau_1}+m_T^{\tau_2}$ $>$ 450 GeV & 2.4 & 0.4 & 0 & 0 & 0 & 2.10 & 1.99 \\
    \hline\hline
    Projections ${\cal L}$ = 3 ab$^{-1}$ & 7.2 & 1.2 & 0 & 0 & 0 & 3.64 & 3.15 \\
    \hline\hline
\end{tabular}
    \caption{\it Cut flow of expected signal and background events for a LHC center-of-mass-energy of $\sqrt{s}$ = 14 TeV and a total integrated luminosity of ${\cal L}$ = 1 ab$^{-1}$, considering a hadronic $\tau$-tagging efficiency of 90\%. Selection cuts from Eq.~\eqref{SRdefinitions} and `MET cuts' from Eq.~\eqref{pTcuts}. Significances from Eqs.~(\ref{statS}) and~(\ref{systS}), the latter with a background systematic uncertainty of 30\%. A QCD multijet estimation of 0.3 events \cite{Aaboud:2018mna} is included in the significances of the last step (accordingly, 0.9 for the projection).}
    \label{cutflowtau90}
\end{table}

\vspace{3mm}
The results of our search strategy, applied step by step on the signal and background events simulated at 14 TeV and a luminosity of 1 ab$^{-1}$, are shown through the cut flow of Table~\ref{cutflowtau90}, for which we have considered a hadronic $\tau$-tagging efficiency of 90\%. The selection cuts leave us with 25\% of the total signal events, but they are really efficient in reducing the backgrounds: $t\bar{t}+2j$ (inc.) and $t\bar{t}+X$ decrease by almost 3 orders of magnitude, and the Diboson and $V$+jets backgrounds are reduced by two orders of magnitude. 
The `MET cuts' cause us to lose only two signal events, leaving us with more than 60\% after the selection cuts, while 25\% of the $t\bar{t}+2j$ (inc.), $t\bar{t}+X$ and Diboson backgrounds survive, and only 3\% of $V$+jets. The $m_\text{eff}$ cut reduces all remaining background events by one order of magnitude, while only eliminating one signal event. 
Finally, the $m_T^{\tau_1}+m_T^{\tau_2}$ cut hardly affects the signal and completely kills all the surviving background events except 0.4 of $t \bar t+2j$ (inc.). 
As we anticipated, we include a QCD multijet estimation (0.3 events) in the significances at the end of the search strategy.
With this cut flow we obtain signal significances, both statistical and with systematic uncertainties of 30\%, around 2 standard deviations. The projections for a total integrated luminosity of 3 ab$^{-1}$ are actually promising, obtaining a statistical signal significance near 4$\sigma$. When considering a conservative 30\% systematic uncertainties in the background, the significance is reduced, but retains a value above the level of evidence (3$\sigma$).

On the other hand, it is important to note that if we consider a more moderate working point for the $\tau$-tagging efficiency of $\epsilon_\tau$=75\%, these results are hardly altered. The signal significances for a luminosity of 1 ab$^{-1}$ would be $\cal{S}_\text{sta}$ = 1.75 and $\cal{S}_\text{sys}$ = 1.69. The corresponding projections for ${\cal L}$ = 3 ab$^{-1}$ would reach values of $\cal{S}_\text{sta}$ = 3.04 and $\cal{S}_\text{sys}$ = 2.73, both signal significances remaining close to the evidence level again.

Furthermore, the robustness of our results in Table~\ref{cutflowtau90} resides in the estimation of the QCD multijet background (it is negligible in the ATLAS search~\cite{Aaboud:2018mna}) and the conservative 30\% systematic uncertainties in the backgrounds for the last HL-LHC upgrade~\cite{HLLHCsys}. 
Hence, following our search strategy, if we neglect the QCD multijet the resulting significance with 30\% systematics for ${\cal L}$ = 1 ab$^{-1}$ (3 ab$^{-1}$) is 2.37$\sigma$ (3.84$\sigma$). 
On the other hand, keeping the estimated 0.3 events for the multijet but considering 20\% of systematics uncertainties we obtain ${\cal S}_\text{sys}$= 2.05$\sigma$ (3.39$\sigma$) for ${\cal L}$ = 1 ab$^{-1}$ (3 ab$^{-1}$).
The most optimistic case, QCD multijet under control and reducing the systematics to 10\%, yields to the promising results ${\cal S}_\text{sys}$ = 2.46$\sigma$ (4.22$\sigma$) for ${\cal L}$ = 1 ab$^{-1}$ (3 ab$^{-1}$).

Finally, the main message of our proof-of-concept collider analysis is presenting a new signal, beyond the usual simplified models, in which the dark matter candidate is Higgsino-like and can be produced from a gluino with a bino-like neutralino as intermediate state. Then the SM-like Higgs boson is also produced and it is tagged by a $\tau$-lepton pair. 
The resulting experimental signature was already analyzed by ATLAS and CMS, however their corresponding SUSY chains involve sleptons as light degrees of freedom whereas they are decoupled in our proposed spectrum.
This SUSY spectrum was analyzed in~\cite{Arganda:2021lpg} with $b$-jets in the final state but it is very interesting to explore the sensitivity to gluinos and Higgsino-like dark matter with $\tau$ leptons in the final state.
The results obtained by the particular SUSY scenario studied here, with conservative background analysis, encourage the development of dedicated interpretations by the LHC experiments.


\section{Conclusions}
\label{conclus}

In this work we propose a new signal at the LHC based on the production of a pair of gluinos, which produce an asymmetric decay chain. One decays directly into the LSP (Higgsino-like) plus jets, while the other decays into jets plus a bino, which in turn decays into the LSP and the Higgs boson. Considering the decay of the Higgs boson into a pair of $\tau$ leptons, the experimental signature consists of 4 jets, 2 $\tau$ and a large amount of missing transverse energy. We identify as the most problematic SM background $t \bar t$+jets and treat in a conservative way the QCD multijet background. Our cut-based search strategy allows us to obtain signal significances, with 30\% systematic uncertainties in the backgrounds, of 2 standard deviations for a center-of-mass energy of 14 TeV and a total integrated luminosity of 1 ab$^{-1}$. The projections for 3 ab$^{-1}$ are promising, increasing this significance to values above the level of evidence. If we consider that for this luminosity the multijet background is under control and negligible, the significance would be slightly lower than 4$\sigma$. Finally, in a more optimistic scenario, if the statistical uncertainty in the background were reduced to 10\%, we would obtain significances greater than 4 standard deviations.

\section*{Acknowledgments}
The work of EA and RM is partially supported by the ``Atracci\'on de Talento'' program (Modalidad 1) of the Comunidad de Madrid (Spain) under the grant number 2019-T1/TIC-14019 and by the Spanish Research Agency (Agencia Estatal de Investigaci\'on) through the grant IFT Centro de Excelencia Severo Ochoa SEV-2016-0597 (EA, RM). The work of EA is also partially supported by CONICET and ANPCyT under projects PICT 2016-0164, PICT 2017-2751, and PICT 2017-2765. The work of AD was partially supported by the National Science Foundation under grant PHY-1820860. The work of MQ is partly supported by Spanish MINEICO under Grant FPA2017-88915-P, by the Catalan Government under Grant 2017SGR1069, and by Severo Ochoa Excellence Program of MINEICO under Grant SEV-2016-0588. IFAE is partially funded by the CERCA program of the Generalitat de Catalunya.

\bibliographystyle{JHEP}
\bibliography{lit}

\end{document}